\documentclass[aps,prl,twocolumn,superscriptaddress,preprintnumbers,floatfix,10pt,nofootinbib]{revtex4-1}
\pdfoutput=1
\usepackage{hyperref}
\usepackage{amsmath}
\usepackage{amsfonts}
\usepackage{slashed}
\usepackage{amssymb}
\usepackage{bm}
\usepackage{bbold}
\usepackage{graphicx}
\usepackage[small]{subfigure}
\usepackage{xcolor}

\newcommand{\refcite}[1]{Ref.~\cite{#1}}
\newcommand{\refscite}[1]{Refs.~\cite{#1}}
\newcommand{\Refcite}[1]{Ref.~\cite{#1}}

\newcommand{\eq}[1]{Eq.~\eqref{eq:#1}}
\newcommand{\eqs}[2]{Eqs.~\eqref{eq:#1} and \eqref{eq:#2}}
\newcommand{\Eq}[1]{Eq.~\eqref{eq:#1}}

\newcommand{\fig}[1]{Fig.~\ref{fig:#1}}

\newcommand{\nn}{\nonumber}

\newcommand{\abs}[1]{\lvert#1\rvert}

\newcommand{\ord}[1]{\mathcal{O}(#1)}

\newcommand{\ORd}[1]{\mathcal{O}\Bigl(#1\Bigr)}

\newcommand{\exv}[1]{\langle#1\rangle}

\newcommand{\EXv}[1]{\Big\langle#1\Big\rangle}

\newcommand{\mae}[3]{\langle#1\lvert#2\rvert#3\rangle}

\newcommand{\ket}[1]{\lvert#1\rangle}

\newcommand{\bra}[1]{\langle#1\rvert}

\newcommand{\Braket}[2]{\big\langle#1\big\vert#2\big\rangle}

\newcommand{\tr}{\operatorname{tr}}
\newcommand{\Tr}{\operatorname{Tr}}

\newcommand{\df}{\mathrm{d}}
\newcommand{\id}{\mathbb{1}}

\renewcommand{\vec}[1]{\bm{#1}}

\newcommand{\la}{\lambda}
\newcommand{\w}{\omega}

\newcommand{\cC}{\mathcal{C}}

\newcommand{\cH}{\mathcal{H}}

\newcommand{\cL}{\mathcal{L}}

\newcommand{\cO}{\mathcal{O}}

\newcommand{\lqcd}{\Lambda_\mathrm{QCD}}
\newcommand{\as}{\alpha_{s}}

\newcommand{\cut}{\mathrm{cut}}

\newcommand{\GeV}{\,\mathrm{GeV}}

\allowdisplaybreaks[2]

\providecommand{\sectionPaper}[1]{\section{\boldmath #1}}
\providecommand{\headingAcknowledgments}{\vspace{0.4em}\paragraph{Acknowledgments:}}
\providecommand{\whereIsTheSupplement}{\hyperref[supplement]{Supplemental Material}}

\begin{document}


\preprint{\vbox{\hbox{MIT-CTP 5818}}}
\preprint{\vbox{\hbox{Nikhef 2024-020}}}
\preprint{\vbox{\hbox{DESY-25-034}}}
\title{
Towards a Quantum Information Theory of Hadronization:
\\
Dihadron Fragmentation and Neutral Polarization in Heavy Baryons
}

\author{Rebecca von Kuk}%
\email{rebecca.von.kuk@desy.de}%
\affiliation{Deutsches Elektronen-Synchrotron DESY, Notkestr. 85, 22607 Hamburg, Germany}%

\author{Kyle Lee}%
\email{kylel@mit.edu}%
\affiliation{Center for Theoretical Physics -- a Leinweber Institute, Massachusetts Institute of Technology, Cambridge, MA 02139, USA}%

\author{Johannes K. L. Michel}%
\email{j.k.l.michel@uva.nl}%
\affiliation{Institute for Theoretical Physics Amsterdam and Delta Institute for Theoretical Physics, University of Amsterdam, Science Park 904, 1098 XH Amsterdam, The Netherlands}%
\affiliation{Nikhef, Theory Group, Science Park 105, 1098 XG, Amsterdam, The Netherlands}%

\author{Zhiquan Sun}%
\email{zqsun@mit.edu}%
\affiliation{Center for Theoretical Physics -- a Leinweber Institute, Massachusetts Institute of Technology, Cambridge, MA 02139, USA}%

\date{June 25, 2025}

\begin{abstract}
We pioneer the application of quantum information theory
to experimentally distinguish between classes of hadronization models.
We adapt the CHSH inequality
to the fragmentation of a single parton to hadron pairs,
a violation of which would rule out classical
dynamics of hadronization altogether.
Furthermore, we apply and extend
the theory of quantum contextuality
and local quantum systems
to the neutral polarization of a single spin-1 hadronic system,
namely the light constituents
of excited Sigma baryons $\Sigma^{*}_{c,b}$
formed in the fragmentation of heavy quarks.

\end{abstract}

\maketitle

\sectionPaper{Introduction}
\label{sec:intro}

Every single time that quarks and gluons are produced in particle collisions,
they combine back into color-neutral bound states like the proton or the pion,
collectively called hadrons.
Calculating this nonperturbative, strongly coupled phenomenon called hadronization
from first principles using the Lagrangian of Quantum Chromodynamics (QCD)
remains an open problem,
restricting us to phenomenological models. 
Perhaps surprisingly, and despite the underlying dynamics of quarks and gluons
clearly being quantum mechanical (QM) in general,
classical stochastic models of hadronization~\cite{Field:1977fa, Field:1989uq, Andersson:1983ia, Webber:1983if}
as e.g.\ implemented in multi-purpose Monte-Carlo (MC) generators~\cite{Bierlich:2022pfr, Bewick:2023tfi, Sherpa:2024mfk}
are vastly successful in this regard,
notably for open, semi-inclusive hadronization observables.
A macroscopic, statistical-mechanics approach especially
succeeds in the large-multiplicity environment of heavy-ion collisions~\cite{Hagedorn:1965st, Becattini:1997rv, Andronic:2003zv, Chojnacki:2011hb}.
In this letter we aim to identify experimental measurement outcomes
that
indicate a departure from classical expectations and
-- if observed -- can only
be described by a full QM treatment of hadronization,
see e.g.\ \refscite{Metz:2002iz, Kerbizi:2021pzn, Kerbizi:2023cde, Masouminia:2023zhb}.
This, to the reader, will be a familiar task:
Asking whether \emph{local reality itself} is quantum
in the form of Bell's inequality~\cite{Bell:1964kc}
lies at the root of today's rich field of quantum information (QI) theory.
We will therefore adapt and extend QI concepts
to hadronization physics
to ask, broadly speaking: \emph{Is hadronization quantum?}

Our proposal is part of a larger fruitful effort to bring QI theory
to bear on questions in particle physics~\cite{Tornqvist:1980af, BESIII:2018cnd, Neill:2018uqw,
Afik:2020onf, Takubo:2021sdk, Severi:2021cnj, Afik:2022kwm, Aoude:2022imd,
Afik:2022dgh, Dong:2023xiw, Morales:2023gow, ATLAS:2023fsd,
Barr:2024djo, Wu:2024mtj, Morales:2024jhj, Chen:2024syv,
Kowalska:2024kbs, Fabbrichesi:2024wcd, Bernal:2024xhm, CMS:2024pts, Afik:2025grr, Guo:2024jch, White:2024nuc, Wu:2024asu, Fabbrichesi:2024rec, Demina:2024dst, LoChiatto:2024dmx, Ruzi:2024cbt, Morales:2024gro, Grossi:2024jae, Cheng:2024rxi, Fabbrichesi:2025ywl, Cheng:2025cuv,
Fabbrichesi:2025ifv, Fabbrichesi:2025aqp}, but distinct
in that most of the existing work,
if not attempting to \emph{explain} the Standard Model (SM)~\cite{Cervera-Lierta:2017tdt, Mulders:2018ion, Thaler:2024anb},
aims to probe quantum foundations
either at the highest energy scales or in exclusive hadronic processes,
essentially asserting the quantum nature
of the perturbative SM Lagrangian
or hadronic form factors, respectively.
Hadronization
then either serves as an analyzer for the perturbative density matrix~\cite{Aidala:2021pvc, Afik:2025grr, Cheng:2025cuv}
or, possibly, as a source of decoherence~\cite{Gong:2021bcp, Datta:2024hpn}
acting on the entangled state created during the perturbative scattering.
By contrast, our proposal directly concerns
the nature of hadronization itself,
and we appeal to principles of effective field theory and
factorization theorems
to isolate the nonperturbative behavior under study
from the (in our view, most likely quantum, and calculable) perturbative scattering
and the self-analyzing hadronic decays.
Our work is also distinct from proposals
to \emph{simulate} hadronization on quantum circuits~\cite{Florio:2023dke, Bauer:2023qgm, Florio:2024aix, Li:2024nod, Lee:2024jnt, Bauer:2025nzf}.

\sectionPaper{Classical hadronization models as ``hidden'' variables}
\label{sec:classical_models_as_local_hidden_vars}

In the picture we are developing,
Bell's ``hidden variables'' $\la$
take on concrete meaning
as a point in the phase space of a stochastic hadronization model
with an intermediate classical state $p(\la)$, i.e., a probability density.
Of course, in this case the variables are not really hidden:
In an analytic model, one may calculate their statistical distribution;
if the model is implicitly defined by a stochastic algorithm,
e.g.\ a Markov process acting on an ensemble
of strings or preconfined clusters~\cite{Amati:1979fg, Andersson:1983ia, Webber:1983if},
one can print out and bin the intermediate state to record $p(\la)$.
One then shows that for large classes of such models,
suitable correlation observables $\cC$
have a bounded expectation value $\langle \cC \rangle_{p(\la)} \leq \cC_\mathrm{max}$,
using basic properties of the integral measure
in any number of classical phase-space dimensions,
while QM expectation values can exceed the bound.
Note that $p(\la)$ and the conditional measurement probabilities
of classical hadronization models must obey
the spacetime symmetries of QCD (Lorentz covariance and parity),
a constraint that is typically absent in tests of quantum foundations,
but which we will make productive use of in this letter.

To fix ideas,
consider the fragmentation $i \to h_1 h_2 X$
of an unpolarized parton $i = q, g$
into identified hadrons $h_{1,2}$.
If $h_{1,2}$ are spin-$1/2$ baryons,
their most general spin-density matrix
compatible with the symmetries of QCD reads
\begin{align} \label{eq:rho_diff}
\rho
= \frac{\id}{4}
+ \tilde{D}_{LL} \, \hat{S}_{1,L} \otimes \hat{S}_{2,L}
+ \tilde{D}_{TT} \, \hat{\mathbf{S}}_{1,T} \otimes \hat{\mathbf{S}}_{2,T}
\,,\end{align}
where $\hbar = 1$ and $\hat{S}_{j,L}$ ($\hat{\mathbf{S}}_{j,T}$)
is the longitudinal (transverse) component of the spin operator
of hadron $h_{j}$ with respect to the fragmentation axis $\mathbf{z}$
pointing back to the hard collision in the hadron's rest frame.
(In the last term, a contraction between vector indices is understood.)
\Eq{rho_diff} is the polarization state of the system
at long distances of order the
hadron lifetimes, $t_\mathrm{decay} \sim 1/\Gamma_h$,
where a QM description of the system
is appropriate for the exclusive hadronic decays~\cite{supplement}.
By contrast, the nonperturbative hadronization physics
are encoded in normalized dihadron fragmentation functions (DiFFs)~\cite{Metz:2016swz,
Pitonyak:2023gjx, Rogers:2024nhb, Pitonyak:2025lin}
$\tilde{D}_{LL}$~\cite{Huang:2024awn}
and $\tilde{D}_{TT}$,
which are determined from experiments
or computed in a model.
In the latter case, we interpret the DiFFs as matching coefficients
that are fixed by computing spin expectation values
$T_{k\ell}/4 \equiv \exv{S_{1,k} \otimes S_{2,\ell}}$
in a (potentially classical) hadronization model
at the scale $t_\mathrm{confine} \sim 1/\lqcd \ll t_\mathrm{decay}$
where the hadrons form and decouple.
We also have
$t_\mathrm{confine} \gg t_\mathrm{hard}$,
where $t_\mathrm{hard}$
is the very short time scale of the perturbative hard scattering.
This separation of all scales allows us to specifically probe the quantum
nature of hadronization and, while abstract, mirrors
the workflow of MC generators.

\begin{figure}[b]
\includegraphics[width=0.5\textwidth]{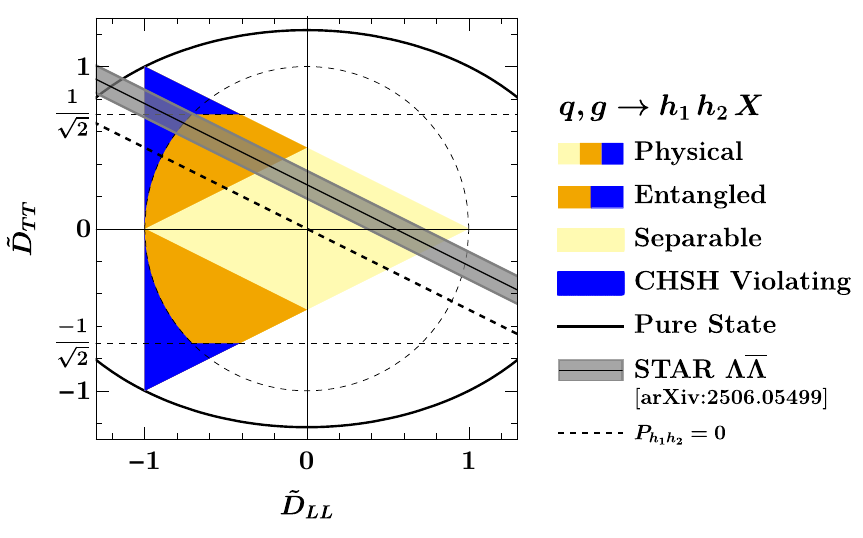}
\caption{
QI characterization of parton fragmentation
to two nearby polarized hadrons.
An observation of dihadron fragmentation functions taking values
in the blue region is inconsistent with classical models of hadronization.
For $\tilde{D}_{LL} = -1$ and $\tilde{D}_{TT} = -1$ ($+1$),
the hadrons are produced in a pure singlet (triplet) Bell state.
For the experimental data, see \refcite{STAR:2025njp, supplement}.
\label{fig:light_diff}
}
\end{figure}

The DiFFs completely characterize the system, see \fig{light_diff}:
Positivity requires $\tilde{D}_{LL} \geq -1$
and $\tilde{D}_{LL} + 2 \abs{\tilde{D}_{TT}} \leq 1$.
By the positive-partial-transpose (PPT) criterion~\cite{Horodecki:1996nc},
the hadrons are entangled
if and only if $\abs{\tilde{D}_{TT}} > (\tilde{D}_{LL} - 1)/2$.
Most interestingly, we can consider the class of fully classical hadronization models
where the (spin) properties of individual final-state hadrons
are separately determined by a common underlying classical state $\la$,
\begin{align} \label{eq:fully_classical_model}
\exv{S_{1,k} \otimes S_{2,\ell}} = \int \! \df \la \, p(\la) \, S_{1,k}(\la) \, S_{2,\ell}(\la)
\,.\end{align}
For these models, the following optimal~\cite{Horodecki:1995nsk}
Clauser-Horne-Shimony-Holt (CHSH) inequality~\cite{PhysRevLett.23.880} holds,
\begin{align} \label{eq:chsh_diff}
\tilde{D}_{TT}^2 + \max \{\tilde{D}_{LL}^2, \tilde{D}_{TT}^2\} \stackrel{\text{\eq{fully_classical_model}}}{\leq} 1
\,.\end{align}
This is the key result for this example process, illustrating our proposal:
If \eq{chsh_diff} is found to be violated in data,
the observation cannot be explained by any fully classical hadronization model
of the form \eq{fully_classical_model}.%
\footnote{
By comparison, the observed Bose-Einstein enhancement of nearby identical
hadrons~\cite{Andersson:1985qn, Artru:1986vc, OPAL:1992thy}
simply follows from (anti)symmetrizing the wave function in momentum space.
Doing so on top of a classical model
leaves the normalized \eq{rho_diff} unaffected,
making \eq{chsh_diff} a deeper test of genuine QM dynamics.
}
The derivation exactly follows Bell's concise proof of the CHSH inequality in \refcite{Bell:2004gpx:chapter4},
with one crucial difference:
In our case, the ``locality'' assumption in \eq{fully_classical_model} is
a property that can be checked
for any given hadronization model (without computation)
by inspecting its dynamical rules and coupling to hadron spins,
and not a property of
a Bell-type local hidden-variable theory (LHVT) of Nature.
The CHSH test we propose here therefore does not test quantum nonlocality
in the usual way, but instead \emph{uses the identical mathematics}
to test whether the blackbox Hamiltonian of hadronization
admits a classical description.%
\footnote{
Another assumption is
that Gedanken measurements
$\exv{(\vec{e}_{\alpha} \cdot \vec{S}_1) \otimes (\vec{S}_2 \cdot \vec{e}_{\beta})}$
at Bell angles $\alpha, \beta$ are indeed predicted
to evaluate to $\vec{e}_{\alpha} \cdot (\mathsf{T}/4) \cdot \vec{e}_{\beta}$,
where $\mathsf{T} = (T_{k\ell})$.
In our case, this property is inherited from
the interpolating partonic spin operators defined in the underlying QCD theory.
Therefore, the counterexamples of \refscite{kasday1970einstein, Abel:1992kz, Li:2024luk}
(\emph{nota bene} constructed to argue against the viability
of testing LHVTs \emph{of Nature} at colliders)
do not apply.
}

\sectionPaper{Quantum contextuality and local spin-1 systems}
\label{sec:quantum_contextuality_and_local_spin_1}

A fundamental aspect of QM systems
is their (statistical) \emph{contextuality}~\cite{Budroni:2021rmt},
the idea that a measurement's outcome
may depend on which other properties are being measured.
To adapt the notion of statistical contextuality to classical hadronization models,
consider
a very simple hadronizing system:
the production of excited heavy spin-$3/2$ baryons $\Sigma_Q^*$
during the fragmentation $Q \to \Sigma_Q^* X$ of beauty or charm quarks, $Q = b,c$.
The light hadron constituents
that bind to the static heavy quark carry total angular momentum $j=1$,
forming a qutrit.
By parity and rotational invariance about the fragmentation axis $\mathbf{z}$,
their quantum polarization state $\rho$
at $t_\mathrm{decay} \sim 1/\Gamma_{\Sigma_Q^*} \gg t_\mathrm{confine}$
is governed by a single nonperturbative Falk-Peskin parameter $w_1$~\cite{Falk:1993rf},
\begin{align} \label{eq:rho_fp}
\rho
= \frac{w_1}{2} \bigl( \ket{1_z}\bra{1_z} + \ket{-1_z}\bra{-1_z} \bigr)
+ \bigl( 1 - w_1 \bigr) \ket{0_z}\bra{0_z}
\,,\end{align}
where the $\ket{m_z}$ are the eigenstates of $\hat{J}_z$.
In a classical hadronization model
at $t_\mathrm{confine} \ll t_\mathrm{decay}$
with intermediate state $p(\la)$,
the probability of measuring
the squared angular momentum
$J_e^2 = (\vec{e} \cdot \vec{J})^2$
along a unit vector $\vec{e}$ to be $x \in \{0,1\}$ reads
\begin{align} \label{eq:P_x_e_hidden_var_model}
P(x|\vec{e}) = \int \df\lambda\, p(\lambda)\, P(x|\lambda,\vec{e}) \,, \quad \langle J_e^2 \rangle = P(1|\vec{e})
\end{align}
and is subject to the constraint
$\langle \vec{J}^2 \rangle = 2$
inherited from the interpolating partonic spin operator,
i.e., $\langle J_a^2 + J_b^2 + J_c^2 \rangle = 2$
for any orthonormal triad $\{\vec{a},\vec{b},\vec{c}\}$.
By Gleason's theorem~\cite{Gleason:1957xxx, Budroni:2021rmt},
parity, and Lorentz covariance,
$P(1 | \vec{e})$ uniquely coincides
with the single-parameter form of $\exv{J_e^2}$
as computed from \eq{rho_fp} for any $\vec{e}$.
All single-point spin measurements are thus determined
by $P(1 | \vec{z}) = \exv{J_z^2} = w_1$,
which
again
serves as a matching coefficient connecting
the classical (or quantum-mechanical) model at $t_\mathrm{confine}$
to the quantum state at $t_\mathrm{decay}$.

\begin{figure}[b]
\includegraphics[width=0.5\textwidth]{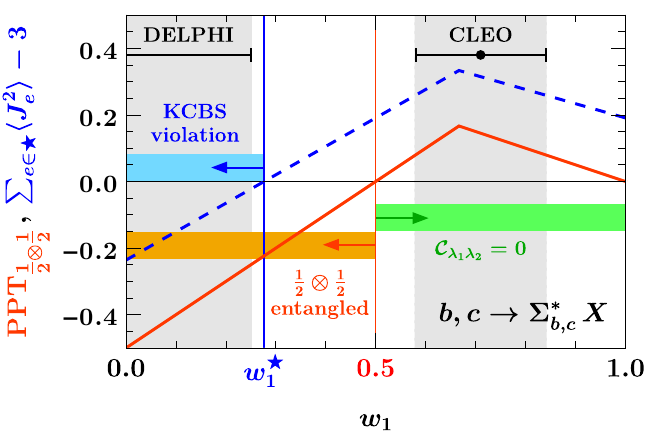}
\caption{
QI characterization of the Falk-Peskin parameter $w_1$
for $Q \to \Sigma_{Q}^{*} X$ in terms of the KCBS inequality (blue),
entanglement in a bipartite QM fragmentation model (orange),
and classical correlation strength in a stochastic model (green).
Experimental data are from \refscite{Feindt:1995qm, CLEO:1996czm},
\label{fig:qi_interpretations_falk_peskin_param}
}
\end{figure}

In this context, \emph{noncontextuality} of a hadronization model
is the statement that in addition to \eq{P_x_e_hidden_var_model},
the model predicts certain joint probabilities
\begin{align} \label{eq:P_x1_xN_e1_eN}
P(x_1, \dots, x_N | \vec{e}_1, \dots, \vec{e}_N)
\end{align}
through a joint conditional measurement probability at given $\la$.
Of particular interest are
$\{\vec{e}_i\}$ arranged in connected Kochen-Specker (KS) graphs~\cite{KS:1967}
where $\vec{e}_i \perp \vec{e}_j$ for all edges in the graph.
The first nontrivial case of a simple cyclic graph
is the Klyachko-Can-Binicioglu-Shumovsky (KCBS) configuration $\bigstar = \{ \vec{e}_i \}$~\cite{Klyachko:2008zz}
of $N = 5$ unit vectors $\vec{e}_i \perp \vec{e}_{i + 1 \!\!\! \mod 5}$
arranged around a symmetry axis $\vec{r} = \vec{z}$,
i.e., they form a regular pentagram in the plane
of fixed $\vec{r} \cdot \vec{e}_i = 1/\sqrt[4]{5}$.
Crucially,
even though the additional measurements
in \eq{P_x1_xN_e1_eN} are counterfactual in our case,
the \emph{existence} of the joint distribution
does leave an imprint on experimentally observable distributions.
This is the statement of the KCBS inequality~\cite{Klyachko:2008zz},
which involves two-point correlations.
Using $\exv{\vec{J}^2} = 2$,
one readily shows (also in the classical case)
that \eq{P_x1_xN_e1_eN} exists for $\bigstar$
if and only if the single-point marginal probabilities satisfy
$\sum_{\vec{e} \in \bigstar} \exv{J_e^2} \geq 3$.
Importantly, these latter single-point measurements
\emph{are} accessible experimentally
and determined by $w_1$.
We thus find a KCBS bound for the Falk-Peskin parameter,
see \fig{qi_interpretations_falk_peskin_param},
\begin{align} \label{eq:w1_kcbs_bound}
w_1 \geq w_1^\bigstar \equiv \frac{1}{2} - \frac{1}{2 \sqrt{5}} \approx 0.276393
\,,\end{align}
which is satisfied if and only if \eq{P_x1_xN_e1_eN} exists for $\bigstar$.
We have verified that the constraint from $\bigstar$ with $N = 5$
is the strongest constraint on $w_1$
from any simple $N$-cycle symmetric around $\vec{z}$.
(In practice, we also choose $\vec{r}$
such that $\sum_{e \in \bigstar} \exv{J_e^2}$ is minimized.)

It is important to realize that
classical models do not have to be noncontextual,
cf.\ \eq{w1_hidden_vector_model}.
Instead, the degree of contextuality, like \eq{fully_classical_model}, is a property
of the hadronization model that can be assessed
by inspecting its dynamics and coupling to the final-state hadron spin.
(The intuition of the KCBS-noncontextual limit being ``more classical''
is valuable nevertheless, cf.\ the case of a thermal system
at high temperatures, where classical and quantum dynamics become indistinguishable
and isotropic, $w_1 \to 2/3 > w_1^\bigstar$.)
Conversely,
the degree to which hadronization models can be coupled to hadron spin
in a noncontextual way
has fundamental limits.
The tightest such constraint for $j = 1$ arises from the Yu-Oh configuration~\cite{PhysRevLett.108.030402},
a KS graph with $N = 13$ containing several cycles,
whose associated state-independent noncontextuality inequality
when combined with $\exv{\vec{J}^2} = 2$ rules out
the existence of \eq{P_x1_xN_e1_eN} for this case altogether.

\sectionPaper{Classical correlation strength}
\label{sec:classical_models}

To explore further interpretations of $w_1$,
consider the ``hidden'' state
of a classical hadronization model
that simply consists of a unit vector $\vec{\la}$ in the hadron rest frame.
Evaluating $P(1 | \vec{z})$ in this model,
we find
\begin{align} \label{eq:w1_hidden_vector_model}
w_1 &= \frac{2}{3}
+ r \, \Delta_2
\,, \quad
\Delta_2 \equiv
\int \! \df c_\theta \,
p(c_\theta) \,
\Bigl(
   c_\theta^2
   - \frac{1}{3}
\Bigr)
\,,\end{align}
where
$c_{\theta} \equiv \vec{z} \cdot \vec{\la}$
and
$r$ is a free parameter governing the strength
of the coupling between $J_e^2$ and $\vec{\la}$
in the Lorentz-invariant conditional measurement probability
$P(1 | \vec{\la}, \vec{e})
= 2/3 + r [(\vec{e} \cdot \vec{\la})^2 - 1/3]$.
Since $0 \leq P(1 | \vec{\la}, \vec{e}) \leq 1$,
we have $-1 \leq r \leq \tfrac{1}{2}$.
If $\vec{\la}$ indicates e.g.\ the average orientation
of classical diquark angular momenta near the heavy quark,
one physically expects a positive coupling $r > 0$
to the observed angular momentum,
in which case $w_1 \geq 1/2$ since $\Delta_2 \geq - 1/3$,
exhibiting KCBS noncontextuality.
However, nothing a priori forbids $r < 0$.
Since $\Delta_2 \leq 2/3$,
we can have
$w_1 \to 0$ in the extreme limit
$r \to -1$ and
$p(c_\theta) \to \delta(\abs{c_\theta} - 1)/2$.
This is of interest as an example of a classical model
that achieves maximal KCBS violation,
and thus must be contextual.

Another important insight from this limit is
that $w_1$ measures the degree
of \emph{classical} correlation \emph{within} the model.
To make this explicit,
consider a bipartite classical model
consisting of two unit vectors $\{ \vec{\la}_1, \vec{\la}_2 \}$
that
couple to the hadron spin
through
$\vec{\la} = (\vec{\la}_1 + \vec{\la}_2)/ \abs{\vec{\la}_1 + \vec{\la}_2}$.
In this case we can prove a general bound
$w_1 \geq ( 1 - |\cC_{\la_1 \la_2}|)/2$,
see \fig{qi_interpretations_falk_peskin_param},
where
\begin{align} \label{eq:def_C_la1_la2}
\cC_{\la_1 \la_2} \equiv \EXv{\frac{
   2 \la_{1z} \la_{2z} - \vec{\la}_1 \cdot \vec{\la}_2 \, (\la_{1z}^2 + \la_{2z}^2)
}{
   1 - (\vec{\la}_1 \cdot \vec{\la}_2)^{2}
}}_{}
\leq
1
\,,\end{align}
is a nonlinear correlation coefficient
that vanishes e.g.\ if $\vec{\la}_{1,2}$ are independent,
$p(\vec{\la}_1, \vec{\la}_2) = p(\vec{\la}_1) \, p(\vec{\la}_2)$.
Alternatively, we can bound $w_1$ for $r<0$
by linear correlations, leading to
\begin{align} \label{eq:w1_two_hidden_vectors}
w_1 \geq
\min \Bigl\{
\exv{\vec{\la}_{1T} \cdot \vec{\la}_{2T}}
,\,
\frac{2}{3}
\Bigr\}
\,,\end{align}
where $\vec{\la}_{iT}$
are the components of $\vec{\la}_i$ transverse to $\vec{z}$.

\sectionPaper{Bipartite QM models and factorization}
\label{sec:qm_models}

Lastly, we interpret \eq{rho_fp}
in a simple QM model of hadronization
where the hadron is formed from the heavy quark
and two light spin-$1/2$ degrees of freedom,
e.g.\ constituent quarks.
Here it is interesting to ask
whether the light system
is entangled or separable
when the hadron is formed.
To assess this, we convert \eq{rho_fp}
to a two-qubit density matrix
(noting that the singlet state,
which would correspond to the $\Lambda_Q$ baryon,
is not populated at this binding energy),
and again evaluate the PPT criterion~\cite{Horodecki:1996nc}.
The result for the lowest eigenvalue of the partial transpose
is shown in red in \fig{qi_interpretations_falk_peskin_param}:
As expected, an overlap of $\Tr[\rho \, \ket{0_z}\bra{0_z}] = 1 - w_1 > 1/2$
with $\ket{0_z}$, which acquires an interpretation
as a triplet Bell state in this model, implies entanglement
(and eventually leads to KCBS violation~\cite{Klyachko:2008zz}).

Importantly, an observation of entanglement at $w_1 < 1/2$
would imply that no factorization can exist
(in the technical sense of the word in high-energy physics)
in terms of any spin-$1/2$ degrees of freedom at cross-section level.
To see this, compare to the prototypical factorization
of the hadronic tensor for the Drell-Yan process at leading-power,
\begin{align} \label{eq:lp_tmd_fact_dy}
W^{\mu\nu}_\mathrm{DY} \sim \tr[\gamma^\mu \Phi_q \gamma^\nu \Phi_{\bar q}]
\,,\end{align}
where
$\Phi_{q,\bar q}$ are (anti)quark correlation functions
of the (possibly polarized) incoming protons,
$\tr$ denotes a trace over Dirac indices,
and $\mu, \nu$ encode the polarization of the virtual photon.
\Eq{lp_tmd_fact_dy} is manifestly separable with respect
to the product $\cH_q \otimes \cH_{\bar q}$ of the (anti)quark Hilbert spaces~\cite{Boer:2004mv}.
Conversely, the neutral polarization state
$\tfrac{1}{\sqrt{2}} \bigl(
   \ket{\uparrow_q \downarrow_{\bar q}}
 + \ket{\downarrow_q \uparrow_{\bar q}}
\bigr)$,
which is maximally entangled,
cannot have large overlap with \eq{lp_tmd_fact_dy},
and indeed it is well known~\cite{Levelt:1993ac, Boer:2008fr,
Bacchetta:2019qkv, Vladimirov:2021hdn,
Ebert:2021jhy, Gamberg:2022lju, delCastillo:2023rng, Gao:2025xxx}
that neutral photon polarization requires an additional spin-$1$ degree of freedom
(and thus subleading-power factorization).
The above argument provides a novel perspective on this fact from QI theory.

\sectionPaper{TMD heavy-quark fragmentation}

\begin{figure}[b]
\includegraphics[width=0.5\textwidth]{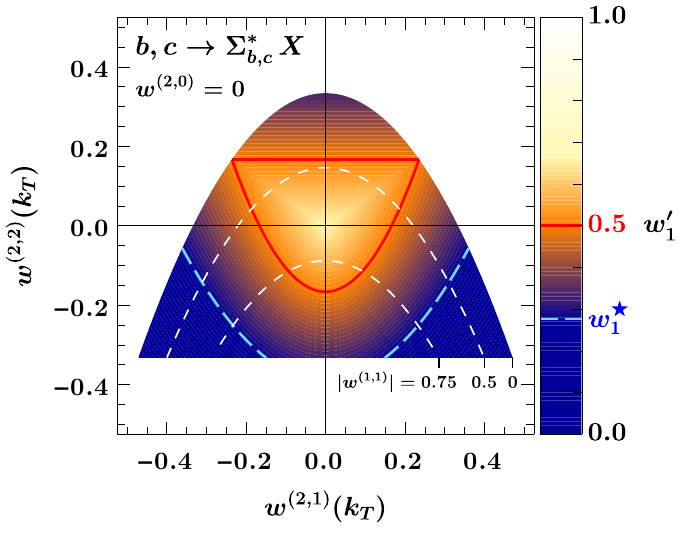}
\caption{
QI characterization of $Q \to \Sigma_{Q}^* X$ with a TMD measurement.
The heatmap indicates the effective longitudinal polarization $w_1'$.
The bounds at $1/2$ and $w_1^\bigstar$ relevant for its QI interpretation
are to be compared to \fig{qi_interpretations_falk_peskin_param}.
\label{fig:qi_interpretations_tmd}
}
\end{figure}

It is experimentally possible~\cite{CLEO:1996czm}
that inclusive fragmentation $Q \to \Sigma_{Q}^* X$
in fact produces an isotropic spin state with $w_1 \approx 2/3$.
It is interesting to ask, therefore,
whether breaking the azimuthal symmetry
by measuring a \emph{transverse} momentum $\vec{k}_T \perp \vec{z}$
provides access to
richer spin-$1$ dynamics.
Here the transverse momentum-dependent (TMD) observable
$\vec{k}_T \gtrsim \lqcd$,
may e.g.\ be
(a)~the recoil of the heavy hadron pair
in $e^+e^- \to \Sigma_{Q}^* H_{\bar{Q}} X$~\cite{vonKuk:2023jfd, Dai:2023rvd, vonKuk:2024uxe, Lee:2025yyy},
(b)~the hadron transverse momentum relative to a jet
containing it~\cite{Kang:2020xyq, Arratia:2020nxw, Lee:2025yyy},
or (c)~the transverse momentum
of an additional fragmentation pion~\cite{Bacchetta:2023njc, Lee:2025yyy}.
In any of these cases, the most general light angular momentum state
compatible with the symmetries reads~\cite{Lee:2025yyy}
\begin{align} \label{eq:rho_tmd}
\rho
&= \frac{\id}{3}
- \tfrac{1}{2} \hat{J}_{y} \, w^{(1,1)}(k_T)
+ \bigl( \tfrac{3}{2} \hat{J}_{z}^2 - \id \bigr) \, w^{(2,0)}(k_T)
\\[0.2em] &
+ \bigl( \hat{J}_x \hat{J}_z + \hat{J}_z \hat{J}_x \bigr) \, w^{(2,1)}(k_T)
+ \bigl( \hat{J}_x^2 - \hat{J}_y^2 \bigr) \, w^{(2,2)}(k_T)
\nn \,,\end{align}
where $k_T = \abs{\vec{k}_T}$,
$\vec{x} = \vec{k}_T/k_T$,
$\vec{y} = \vec{z} \times \vec{x}$,
and the $w^{(N,n)}$ are the nonperturbative coefficients.
Our previous discussion
elegantly carries over if we define
\begin{align} \label{eq:def_w1_prime}
w_1'
\equiv
\exv{J_{z'}^2}
\equiv \min_{\vec{e}} \, \exv{J_e^2}
= \min_{\vec{e}} \, \Tr\bigl[ \rho \, \hat{J}_e^2 \bigr]
\,,\end{align}
i.e., by finding the axis $\vec{z'}$ that maximizes
the neutral polarization rate $1 - w_1'$.
A rotation from $\vec{z}$ to $\vec{z}'$
in fact diagonalizes $\rho$
with
eigenvectors $\ket{m_{z'}}$,
so we can think of $\vec{z'}$ as a novel polarization axis
that is \emph{spontaneously generated} by the fragmentation dynamics.
Here we can make immediate use of $\vec{z}'$
by repeating our previous QI analysis for $w_1'$,
as shown in \fig{qi_interpretations_tmd},
where $w^{(2,0)} = 0$ corresponds to $w_1 = 2/3$
after integrating over $\vec{k}_T$.
The value of $w_1'$ is driven by $w^{(2,1)}$ and $w^{(2,2)}$,
and even for $w_1 = 2/3$
we find physical regions in their parameter space
with nontrivial QI interpretations $w_1' < 1/2$ or $w_1' < w_1^\bigstar$,
respectively.
(Increasing $w^{(1,1)}$ shrinks the physical region,
but leaves $w_1'$ unchanged.)
Notably,
their scaling for $k_T \gg \lqcd$
is model-independently predicted
to be $w^{(N,n)}(k_T) \sim (\lqcd/k_T)^n$~\cite{Lee:2025yyy}.
Therefore if
heavy-quark fragmentation indeed exhibits
``quantum'' behavior
at $k_T \sim \lqcd$ as in \fig{qi_interpretations_tmd},
there would be the exciting possibility
to observe a quantum-to-classical transition as $k_T$ increases.

\sectionPaper{Conclusions and outlook}

In this letter we have initiated the study of
applying tools from quantum information (QI) theory
to the open problem of hadronization to distinguish between entire
classes of classical or quantum-mechanical hadronization models at once.
In \eq{chsh_diff},
we have presented an inequality
for light dihadron fragmentation
that, if violated in data,
would rule out classical hadronization dynamics altogether.
We stress that our goal was not to test quantum foundations themselves:
Rather, we adapted tools from QI theory to characterize
the dynamics of the blackbox Hamiltonian of hadronization,
and appealed to effective field theory to separate it from other
(very likely quantum) dynamics in the problem.
We then adapted the concept of contextuality
to the study of hadronization,
showcased our ideas
using the Falk-Peskin fragmentation parameter $w_1$
in heavy baryon production,
and further interpreted $w_1$
in terms of classical correlation strength
and QM entanglement,
with implications for factorizability.
Incorporating these insights
will impose practical constraints
on future hadronization modeling
and enable cross-pollination with ongoing QI research.

The experimental status of $w_1$ is unclear at present,
see \fig{qi_interpretations_falk_peskin_param}.
Combining $\Sigma_b^{* \pm}$ baryons, DELPHI measured~\cite{Feindt:1995qm}
\begin{align} \label{eq:w1_delphi}
w_{1}^\mathrm{DELPHI} = -0.36 \pm 0.30_\mathrm{stat} \pm 0.30_\mathrm{syst}
\,,\end{align}
which translates to a 95\% CL limit of $w_1 \leq 0.25$
after imposing positivity.
On the other hand, by combining $\Sigma_c^{*++}$ and $\Sigma_c^{*0}$ baryons
(that have the identical light valence content and relative rates),
CLEO found~\cite{CLEO:1996czm},
\begin{align} \label{eq:w1_cleo}
w_{1}^\mathrm{CLEO} = 0.71 \pm 0.13
\,,\end{align}
close to the isotropic limit $w_1 = 2/3$.
This disagreement has, to our knowledge, not been resolved~\cite{Falk:1996nv}.
We therefore strongly encourage a new measurement of $w_1$
using state-of-the-art experimental techniques.
We also encourage experimental tests of spin correlations
in nearby dihadrons~\cite{STAR:2025njp}
and of the TMD generalizations
of $w_1$ that we introduced, all of which possess deep physical interpretations
within the new QI framework of hadronization that we have presented here.

\begin{acknowledgments}
\headingAcknowledgments
We would like to thank Dani{\"e}l Boer, Jordi Tura Brugu\'es, Herbi Dreiner,
Jonas Helsen, Piet Mulders,
Duff Neill,
Marieke Postma,
Iain Stewart, and Jesse Thaler
for fruitful discussion, as well as the members of the Nikhef LHCb Group
and Mick Mulder in particular.
The authors gratefully acknowledge the hospitality
of the Nikhef Theory Group, the Institute for Theoretical Physics at the University of Amsterdam, and the Erwin Schr\"{o}dinger Institute.
K.L. and Z.S. were supported by the Office of Nuclear Physics of the U.S.\
Department of Energy under Contract No.\ DE-SC0011090.
Z.S. was also supported by a fellowship from the MIT Department of Physics.
R.v.K. was supported by the European Research Council (ERC) under the European
Union's Horizon 2020 research and innovation programme (Grant agreement No.
101002090 COLORFREE).
J.M. was supported by the D-ITP consortium, a program of NWO that is funded by the Dutch Ministry of Education, Culture and Science (OCW).
\end{acknowledgments}

\bibliography{references}

\onecolumngrid
\clearpage

\section*{Supplemental material}
\label{supplement}

\setcounter{equation}{1000}
\setcounter{figure}{1000}
\setcounter{table}{1000}

\renewcommand{\theequation}{S\the\numexpr\value{equation}-1000\relax}
\renewcommand{\thefigure}{S\the\numexpr\value{figure}-1000\relax}
\renewcommand{\thetable}{S\the\numexpr\value{table}-1000\relax}

In this supplemental material we collect explicit expressions
for the differential distributions of decay products
stemming from the self-analyzing decays
of the polarized hadron systems discussed in the main text.
These expressions involve the nonperturbative fragmentation coefficients
relevant for the analysis in the main text as free parameters,
and allow for their experimental determination
by fitting them to the observed distributions.

\subsection{Fragmentation to nearby dihadrons}
\label{sec:decay_distribution_dihadrons}

\subsubsection{Factorization and general structure}

Consider the cross section for producing a dihadron pair
with general spin $s_{1,2}$
and magnetic quantum numbers  $-s_1 \leq m_1, m_1' \leq s_1$
and $-s_2 \leq m_2, m_2' \leq s_2$
with respect to the fragmentation axis $\mathbf{z}$.
(In the collinear approximation,
the two hadron rest frames only differ by a boost along the $\mathbf{z}$ direction,
such that its orientation is common.)
We decompose the cross section for an arbitrary spin state
in terms of operators $\hat{\cO}$ on the product Hilbert space as
\begin{align}
\frac{\df \sigma_{m_1 m_1', m_2 m_2'}}{\df z_1 \, \df z_2 \, \df R_T}
= \sum_\cO \frac{\df \sigma_\cO}{\df z_1 \, \df z_2 \, \df R_T} \,
\cO_{m_1 m_1', m_2 m_2'}
\,,\end{align}
where $\cO_{m_1 m_1', m_2 m_2'} = \mae{m_1 m_2}{\hat{\cO}}{m_1' m_2'}$
and $z_1, z_2$ are the longitudinal momentum fractions
of hadrons $h_1$ and $h_2$, respectively.
Here, $R_{T}$ is the magnitude of the transverse component of the difference of hadron momenta relative to their sum.
For pairs of spin-$1/2$ baryons we have
\begin{align}
\{ \cO \} = \Bigl\{
\frac{\id}{4} ,\,
\hat{S}_{1,L} \otimes \hat{S}_{2,L} ,\,
\hat{\mathbf{S}}_{1,T} \otimes \hat{\mathbf{S}}_{2,T}
\Bigr\}
\,.\end{align}
as in \eq{rho_diff}, where the coefficients $\df \sigma_\cO$
in turn are also denoted by $\df \sigma$ (without subscript),
$\df \sigma_{LL}$, and $\df \sigma_{TT}$.
They are given in terms of dihadron fragmentation functions $D^\cO_{i\to h_1 h_2}$ as~\cite{Metz:2016swz,Pitonyak:2023gjx, Rogers:2024nhb, Pitonyak:2025lin,Huang:2024awn}

\begin{align}
\frac{\df \sigma_\cO}{\df z_1 \, \df z_2 \, \df R_T}
= \sum_i \! \int \df y \,
C_i(y,\mu) \,
D^\cO_{i\to h_1 h_2}\left(\frac{z_1}{y},\frac{z_2}{y}, R_T,\mu\right)\,,
\end{align}
where $C_i$ are the usual single-inclusive partonic hard coefficients.
(In the case of hadronic collisions, they also contain the parton distribution functions.)
Additionally, $C_i$ can include fiducial acceptance cuts
on the overall direction of the near-collimated dihadron system;
cuts on individual hadron energies or momenta are special cases
of the bins we consider below.

We are interested in the yield $N_\cO$ of dihadron pairs
produced in a specific spin state $\cO$ and bin of $(z_1, z_2)$,
which is proportional to
\begin{align}
\frac{N_\cO}{\cL} = \sigma_{\cO}
&\equiv \int \! \df z_1 \, \df z_2 \int_0^{R_T^{\rm{cut}}} \! \df R_T \,
\frac{\df \sigma_\cO}{\df z_1 \, \df z_2 \, \df R_T} \,
\Theta(z_1, z_2)
\nn \\
&= \sum_i \int \! \df y  \, \df x_1 \, \df x_2 \, C_i(y, \mu) \, \Theta(yx_1, yx_2) \,
\int_0^{R_T^{\cut}} \! \df R_T \,
D^\cO_{i\to h_1 h_2}\bigl(x_1,x_2, R_T,\mu\bigr)
\,,\end{align}
where $\Theta(z_1, z_2)$ implements the given bin
and the normalization factor $\cL$ is the total luminosity.
By placing an upper cutoff $R_T^\mathrm{cut} \sim \lqcd$
on the relative transverse momentum, we ensure
that spin correlations are genuinely sourced
by the nonperturbative fragmentation dynamics.
The total yield $N$ is given by the $N_\cO$ with $\cO = \id/\Tr[\id]$ the identity
on the product Hilbert space.
The normalized density matrix of the hadron pair then reads
\begin{align}
\rho = \frac{1}{N} \sum_\cO N_\cO \, \cO
\,.\end{align}
For spin-$1/2$ baryons this results in \eq{rho_diff} if we define the shorthand
\begin{align} \label{eq:def_d_ll_final}
\tilde{D}_{LL}
= \frac{\sigma_{LL}}{\sigma}
= \frac{
   \sum_i \int \! \df y\, \df x_1\, \df x_2 \, C_i(y, \mu) \, \Theta(yx_1, yx_2) \,
   \int_0^{R_T^{\cut}} \! \df R_T \,
   D^{LL}_{i\to h_1 h_2}\bigl(x_1,x_2, R_T,\mu\bigr)
}{
   \sum_i \int \! \df y \,  \df x_1\, \df x_2 C_i(y, \mu) \, \Theta(yx_1, yx_2) \,
   \int_0^{R_T^{\cut}} \! \df R_T \,
   D_{i\to h_1 h_2}\bigl(x_1,x_2, R_T,\mu\bigr)
}
\,,\end{align}
and similarly for $\tilde{D}_{TT}$.
We stress that even though the DiFFs enter in a convolution
with the perturbative scattering coefficient in this form,
the strength of the spin correlation is still uniquely
sourced at the low scale, since the DGLAP evolution
and the perturbative corrections are all common.
In particular, using the leading-order perturbative result for the coefficient
$C_i(y, \mu) \propto \delta(1 - y) + \ord{\as}$,
\eq{def_d_ll_final} indeed reduces to a ratio of the (binned) DiFFs themselves.

We point out that while the expression for $\tilde{D}_{LL}$
in \eq{def_d_ll_final} is clearly the most amenable
to experimental tests of \eq{chsh_diff} in the near future,
since it directly relates to the measured cross section at a given collider,
we believe it would be valuable in the long run to test \eq{chsh_diff}
directly at the level of the process and collider-independent ratio
\begin{align} \label{eq:def_d_ll_alternative}
\tilde{D}_{LL,i}\bigl(x_1, x_2, R_T^{\cut}\bigr)
\equiv \frac{
   \int_0^{R_T^{\cut}} \! \df R_T \,
   D^{LL}_{i\to h_1 h_2}\bigl(x_1,x_2, R_T,\mu\bigr)
}{
   \int_0^{R_T^{\cut}} \! \df R_T\,
   D_{i\to h_1 h_2}\bigl(x_1,x_2,R_T,\mu\bigr)
}
\,,\end{align}
with $\tilde{D}_{TT,i}$ again defined in full analogy,
where the right-hand side is evaluated at some low input scale $\mu \sim 2 \GeV$.
(In a more sophisticated setup, one may also define $\tilde{D}_{LL}$ and $\tilde{D}_{TT}$
as ratios of Mellin moments of the singlet and gluon or valence DiFFs
to form exact renormalization-group invariants.)
\Eq{chsh_diff} equally well applies to $\tilde{D}_{LL,i}$ and $\tilde{D}_{TT,i}$
when defined as in \eq{def_d_ll_alternative},
where the normalized density matrix is now directly
computed in terms of the renormalized dihadron fragmentation correlator
at the low scale.
Using \eq{def_d_ll_alternative} in practice requires one to
first determine the underlying DiFFs from a global fit to the available collider data
for the yields $N_\cO$ extracted from decay distributions in many different bins.
Nevertheless, we believe this would be worth the effort it entails,
since one would gain access to the degree of CHSH violation
also as a function of e.g.\ the parton type.

\subsubsection{Decay distribution for nearby hyperon pairs}

Differential distributions for the decay products of the hadrons
are obtained by tracing \eq{rho_diff}
against the (conjugate) decay matrix elements for given hadron helicities
$m_{1,2}$ ($m_{1,2}'$),
which are specific to each hadron and decay.
Here we consider the parity-violating weak decay of hyperons,
a standard example of a self-analyzing decay
commonly used for reconstructing (transverse) hadron polarization,
see e.g.\ \refscite{BESIII:2018cnd, Gong:2021bcp, Gamberg:2021iat, BESIII:2022qax, Zhang:2023ugf, Gao:2024dxl, Wu:2024asu}.

To fix conventions we begin with the decay
of a single, individually polarized (anti)hyperon
\begin{align} \label{eq:hyperon_decays}
h_1 = \Lambda \, \to p \, \pi^-
\,, \qquad
h_2 = \bar \Lambda \, \to \bar p \, \pi^+
\,.\end{align}
The associated decay distributions are standard~\cite{ParticleDataGroup:2020ssz}
and read
\begin{align} \label{eq:distribution_individual_hyperon_decays}
\frac{1}{N} \frac{\df N}{\df \cos \theta_{p} \, \df \varphi_{p}}
= \frac{1}{4\pi} \bigl[
   1 + \alpha_- \, \vec{P} \cdot \vec{n}_{p}
\bigr]
\,, \qquad
\frac{1}{N} \frac{\df N}{\df \cos \theta_{\bar p} \, \df \varphi_{\bar p}}
= \frac{1}{4\pi} \bigl[
   1 + \alpha_+ \, \vec{P} \cdot \vec{n}_{\bar p}
\bigr]
\,,\end{align}
where $\vec{P} = 2 \exv{\hat{\vec{S}}}$ with $\abs{\vec{P}} \leq 1$
is the Bloch vector of the (anti)hyperon,
$\vec{n_p}$ ($\vec{n_{\bar p}}$) is a unit vector
pointing in the direction of the (anti)proton in the rest frame of the parent particle,
$\theta_{p}$ and $\varphi_{p}$ ($\theta_{\bar p}$ and $\varphi_{\bar p}$)
are the polar coordinates of $\vec{n_p}$ ($\vec{n_{\bar p}}$)
in the respective rest frame,
and $\alpha_\mp$ are the hyperon decay parameters.
With the above common sign in front of $\alpha_\pm$ for hyperon and antihyperon,
one has opposite signs for $\alpha_- = 0.732 \pm 0.014$
and $\alpha_+ = -0.758 \pm 0.012$~\cite{ParticleDataGroup:2020ssz}
(and thus $\alpha_- \alpha_+ < 0$),
as required by the approximate CP symmetry.

For the system in \eq{rho_diff},
each individual hyperon is indeed unpolarized on average, $\vec{P} = 0$,
such that the distributions $\df N/(\df \cos \theta_{p} \, \df \varphi_{p})$
and $\df N/(\df \cos \theta_{\bar p} \, \df \varphi_{\bar p})$ are flat.
Nevertheless, the spin correlations encoded in \eq{rho_diff}
can be readily observed from the joint distribution
of their decay products,
\begin{align} \label{eq:joint_distribution_hyperon_pair_decays}
\frac{1}{N} \frac{\df N}{\df \cos \theta_{p} \, \df \varphi_{p} \, \df \cos \theta_{\bar p} \, \df \varphi_{\bar p}}
&= \frac{1}{(4\pi)^2} \bigl[
   1
   + \alpha_- \alpha_+ \, \vec{n}_{p} \cdot \mathsf{T} \cdot \vec{n}_{\bar p}
\bigr]
\nn \\
&= \frac{1}{(4\pi)^2} \bigl[
   1
   + \alpha_- \alpha_+ \, \tilde{D}_{LL} \, \cos \theta_{p} \, \cos \theta_{\bar p}
   + \alpha_- \alpha_+ \, \tilde{D}_{TT} \, \sin \theta_{p} \, \sin \theta_{\bar p} \cos(\varphi_{p} - \varphi_{\bar p})
\bigr]
\,,\end{align}
where the rank-two tensor $\mathsf{T} = (T_{k\ell})
= \operatorname{diag}(\tilde{D}_{TT}, \tilde{D}_{TT}, \tilde{D}_{LL})$
has components $T_{k\ell}/4 = \exv{S_{1,k} \otimes S_{2,\ell}}$.
We note that \eq{joint_distribution_hyperon_pair_decays}
takes the same form for the decay products of same-sign hyperon pairs
(or other exclusive baryonic decays that follow \eq{distribution_individual_hyperon_decays})
after suitably replacing the decay constants.

\paragraph{Comparison to \refcite{STAR:2025njp}:} Very recently,
and after the first preprint version of this manuscript was made public,
the STAR collaboration published a measurement~\cite{STAR:2025njp}
of the differential distribution
\begin{align} \label{eq:joint_distribution_hyperon_pair_decays_star}
\frac{1}{N} \frac{\df N}{\df \cos \theta^*}
= \frac{1}{2} \bigl[ 1 + \alpha_- \alpha_+ \, P_{\Lambda \bar{\Lambda}} \, \cos \theta^*\bigr]
\,,\end{align}
where in our notation $\cos \theta^* = \vec{n}_p \cdot \vec{n}_{\bar p}$
and the coefficient $P_{\Lambda \bar{\Lambda}}$ encodes the strength of the angular correlation
between the decay (anti)protons,
and thus the strength of the spin correlation
between the parent (anti)hyperons.
Projecting our \eq{joint_distribution_hyperon_pair_decays} onto the distribution differential in $\cos \theta^*$,
we find the relation
\begin{align}
3 P_{\Lambda \bar{\Lambda}} = \Tr \mathsf{T} = 2\tilde{D}_{TT} + \tilde{D}_{LL}
\,.\end{align}
The measurement of $P_{\Lambda \bar{\Lambda}}$ in \refcite{STAR:2025njp}, as well as those
for same-sign hyperon pairs obtained in \refcite{STAR:2025njp}, therefore experimentally
constrain a very specific linear combination of dihadron fragmentation functions.
\Refcite{STAR:2025njp} found that for nearby dihadrons,
which as discussed in the main text and the above supplemental material
is the most interesting kinematic region,
the spin correlation $P_{\Lambda \bar{\Lambda}} =0.181\pm 0.035_\mathrm{stat}\pm0.022_\mathrm{sys}$
was near its maximal value of $1/3$
(with some loss expected due to feed-down contributions from higher resonances),
while $P_{\Lambda \Lambda}$ and $P_{\bar{\Lambda} \bar{\Lambda}}$
were still compatible with zero.
These experimental constraints are indicated in \fig{light_diff}
as of the second preprint version of this manuscript.
Notably, either of the experimental results is still compatible
with entanglement or even CHSH violation in the upper left corner,
since those depend on the precise breakdown between $\tilde{D}_{LL}$ and $\tilde{D}_{TT}$
and in particular the sign of $\tilde{D}_{LL}$.
We stress that two caveats apply when interpreting the measurement
of \refcite{STAR:2025njp} in this way; specifically, (a)~further work is needed to ascertain
the validity of the factorization in terms of fragmentation functions,
which might require a more stringent cut on the $p_T$ of the hyperon candidates,
and (b) the equivalence
between the respective reference frames
and angles should be carefully checked.

The analysis in \refcite{STAR:2025njp} was in part motivated by theory considerations
that postulate a rotationally invariant distribution of decay products,
as e.g.\ appropriate for exclusive $\Lambda \bar{\Lambda}$ production,
in which case $\cos \theta^*$ is the only angular observable of interest.
As we have discussed, however, the fragmentation process
necessarily retains a dependence on the fragmentation axis $\hat{z}$,
i.e., the orientation relative to the hard scattering process
that produces the parent parton,
and whose hard scattering products
balance the parton's color charge.
For this reason the polarization tensor $\textsf{T}$ need not be proportional
to the identity, and indeed in general has two nontrivial degrees of freedom
($\tilde{D}_{LL}$ and $\tilde{D}_{TT}$) according to our symmetry analysis
that led to \eq{rho_diff}.
We therefore strongly encourage an extension
of the exciting measurement reported in \refcite{STAR:2025njp}
to the more differential angular distribution
in \eq{joint_distribution_hyperon_pair_decays},
which would break the degeneracy between $\tilde{D}_{LL}$ and $\tilde{D}_{TT}$
and hold the potential of a definitive interpretation
in terms of QI constraints on possible hadronization models.
Alternatively, the degeneracy could at this point also be broken
by a simpler, direct measurement of $\tilde{D}_{LL}$
via the $\cos \theta_{p} \cos \theta_{\bar p}$ correlation in \eq{joint_distribution_hyperon_pair_decays}
in a way that is inclusive over $\phi_{p, \bar p}$, i.e.,
\begin{align}
\frac{1}{N} \frac{\df N}{\df \cos \theta_{p} \, \df \cos \theta_{\bar p}}
&= \frac{1}{4} \bigl[
   1
   + \alpha_- \alpha_+ \, \tilde{D}_{LL} \, \cos \theta_{p} \, \cos \theta_{\bar p}
\bigr]
\,.\end{align}

\subsection{Inclusive fragmentation to excited heavy baryons}
\label{sec:decay_distribution_heavy_incl}

The factorization theorem for the production of a boosted polarized heavy hadron $H$
containing a heavy quark of mass $m_Q$
at strict leading power in $\lqcd/m_Q$
is simply a product~\cite{Mele:1990cw, Jaffe:1993ie, Falk:1993rf},
\begin{align} \label{eq:factorization_boosted_heavy_quark_fragmentation}
\frac{\df \sigma_{HX}^{\cO_\ell}}{\df^3 \vec{P}_H}
= \frac{\df \sigma_{QX}}{\df^3 \vec{P}_Q} \, \chi_{H,\cO_\ell}
\Bigl[ 1 + \ORd{\frac{\lqcd}{m_Q}} + \ORd{\frac{m_Q}{\abs{\vec{P}_H}}} \Bigr]
\,.\end{align}
The matching coefficient $\df \sigma_{QX}$ is the differential cross section
for producing a free unpolarized heavy quark
at the same three-momentum in the center-of-mass frame of the collision, $\vec{P}_Q = \vec{P}_H$.
The nonperturbative dynamics are contained
in the fragmentation coefficient $\chi_{H,\cO_\ell}$.
Note that for hadron colliders, the second set of power corrections
in \eq{factorization_boosted_heavy_quark_fragmentation} scales
as ${m_Q}/{\abs{\vec{P}_T}}$ instead,
where $\vec{P}_T$ is the transverse momentum relative to the beam axis.
As before, $\cO_\ell$ labels a spin operator encoding a generic spin state,
in this case for the total angular momentum of the light constituents $\ell$ of $H$.
The total probability $\chi_H$ for $Q$ to fragment into $H$,
which is the coefficient $\chi_{H,\cO}$
of $\cO = \id_\ell/\Tr[\id_\ell]$, satisfies $\sum_H \chi_H = 1$.
Given the multiplicative form
of \eq{factorization_boosted_heavy_quark_fragmentation},
cuts acting on the heavy hadron kinematics
simply act on the heavy-quark production cross section.
Thus, unlike the light dihadron case, no convolution is required to compute
the total yields for each spin state.
Instead, we have
\begin{align}
\rho
= \frac{1}{N} \sum_{\cO_\ell} N_{\cO_\ell} \, \cO_\ell
= \frac{1}{\chi_H} \sum_{\cO_\ell} \chi_{H,\cO_\ell} \, \cO_\ell
\,,\end{align}
i.e., the normalized spin density matrix of the light constituents
in any given bin of the heavy-hadron kinematics
is directly given by ratios of the nonperturbative coefficients.
The normalized coefficients $\chi_{H,\cO_\ell}/\chi_H$
for a light state with angular momentum $j = 1$ or $j = 3/2$
are in direct correspondence to the usual Falk-Peskin parameters $w_j$.
Specifically, for the set of operators
$\{ \cO_\ell \} = \{ \id/4 ,\, \hat{J}_{z}^2\}$
allowed by the symmetries in the case $j = 1$,
one readily converts the coefficients
to the form involving $w_1$ in \eq{rho_fp}
using $\hat{J}_z^2 = \id - \ket{0_z}\bra{0_z}$.

Experimentally, the Falk-Peskin parameter $w_1$ can be measured
from the pion distribution in the dominant decay $\Sigma_Q^{*} \to \Lambda_Q \pi$.
Consistently working to the leading-order in the heavy-quark expansion,
the decay amplitude is given
by the Isgur-Wise transition matrix element~\cite{Isgur:1991wq},
which is proportional to
\begin{align} \label{eq:isgur_wise_transition}
Y_{1m}(\theta, \varphi) \, \Braket{1 m \, \tfrac{1}{2} k}{ s \, h }
\,.\end{align}
Here $Y_{j m}$ is a spherical harmonic,
$\theta$ ($\varphi$) is the pion polar (azimuthal) angle
relative to the fragmentation axis
in the candidate $\Sigma_Q^{(*)} \to \Lambda_Q \pi$ rest frame,
$s = 1/2$ ($3/2$) and $h$ are the total spin
and spin along the $\vec{z}$ axis of the $\Sigma_Q$ ($\Sigma_Q^*$) baryon,
$k$ is the spin of the final-state $\Lambda_Q$ along the $\vec{z}$ axis,
and $m = h - k$.
We recall the following, very simple form of the fragmentation axis $\vec{z}$
when expressed in terms of the beam three-momenta $\vec{P}_{a,b}$
in the candidate rest frame,
\begin{align}
\vec{z} = - \frac{\vec{P_a} + \vec{P_b}}{\abs{\vec{P_a} + \vec{P_b}}}
\,,\end{align}
where the beams can be protons or leptons.
We now dress \eqs{rho_fp}{isgur_wise_transition}
with the appropriate Clebsch-Gordan coefficients,
average over the helicities of the unpolarized initial-state heavy-quark,
and sum over the helicities of the final-state $\Lambda_Q$
(whose polarization is not reconstructed).
This results in the following angular distributions
for the decay products~\cite{Falk:1993rf},
\begin{align} \label{eq:decay_distribution_fp}
\frac{1}{N_{\Sigma_Q}} \frac{\df N_{\Sigma_Q}}{\df \cos \theta}
= \frac{1}{2}
\,, \qquad
\frac{1}{N_{\Sigma_Q^{*}}} \frac{\df N_{\Sigma_Q^{*}}}{\df \cos \theta}
&= \frac{1}{4} \Bigl[ 1 + 3 \cos^2 \theta - \frac{9}{2} \w_1 \Bigl( \cos^2 \theta - \frac{1}{3} \Bigr) \Bigr]
\nn \\
&= \frac{1}{2} \biggl\{
   1
   - \frac{3}{8} \bigl[ 1 + 3 \cos(2\theta) \bigr] \Bigl( w_1 - \frac{2}{3} \Bigr)
\biggr\}
\,.\end{align}
Here we have performed the integral over the $\varphi$ dependence,
which as expected is flat since no azimuthal direction is preferred.
We assumed that the $\Sigma_Q$ and $\Sigma_Q^*$
form well-separated resonances
$\Gamma_{\Sigma_Q}, \Gamma_{\Sigma_Q^*} \ll m_{\Sigma_Q^*} - m_{\Sigma_Q}$
relative to the mass splitting in the heavy-quark spin symmetry doublet,
which is reasonably well satisfied~\cite{CDF:2011ac, LHCb:2018haf}.
If needed, the decay distributions can readily be generalized
to account for their interference~\cite{Falk:1993rf}, and still can be expressed
purely in terms of $w_1$ in that case.
We note that the continuum $Q \to \Lambda_Q \pi$ background to \eq{decay_distribution_fp}
initiated by a boosted heavy quark
is in fact described by a heavy-light DiFF that is of interest of its own,
and upon expansion in $\lqcd/m_Q \ll 1$
takes the same form as \eq{factorization_boosted_heavy_quark_fragmentation}
with a more differential nonperturbative matrix element $\chi_{\Lambda_Q \pi}$,
\begin{align}
\frac{\df \sigma_{\Lambda_Q \pi X}^{\cO_\ell}}{\df^3 \vec{P}_{\Lambda_Q \pi}^\mathrm{lab} \, \df m_{\Lambda_Q \pi} \, \df \cos \theta}
= \frac{\df \sigma_{QX}}{\df^3 \vec{P}_{Q}^\mathrm{lab}} \, \int \! \df E_\pi \, \df p_\pi^z \,
\chi_{\Lambda_Q \pi}(E_\pi, p_\pi^z) \,
\delta\bigl( m_{\Lambda_Q \pi} - m_{\Lambda_Q} - m_\pi - E_\pi \bigr) \, \delta \Bigl( \cos \theta - \frac{p_\pi^z}{\abs{\vec{p}_{\pi}}} \Bigr)
\,,\end{align}
where $m_{\Lambda_Q \pi}$ is the invariant mass of the candidate pair
and $E_\pi$ and $\vec{p}_\pi$ are the energy and three-momentum
of the pion in the candidate center-of-mass frame, $\vec{p}_\pi = - \vec{p}_{\Lambda_Q}$.
(For clarity, we have made explicit that the approximately identical
kinematics $\vec{P}_{\Lambda_Q \pi}^\mathrm{lab} = \vec{P}_{Q}^\mathrm{lab} + \ord{\lqcd}$
of the candidate pair and the heavy-quark progenitor
are evaluated in the lab frame instead;
in practice this dependence will always be integrated over a certain bin
at the level of the prefactor.)
The dependence of $\chi_{\Lambda_Q \pi}$ on $p_\pi^z$,
which determines the $\cos \theta$ shape of the background,
and the correlation between $p_\pi^z$ and $E_\pi$,
satisfy the obvious on-shell constraints $E_\pi \geq m_\pi$
and $\abs{p_\pi^z} \leq \sqrt{E_\pi^2 - m_\pi^2}$,
but are complicated in general.
Importantly, however, nothing privileges
the point $E_\pi \approx m_{\Sigma_Q^*} - m_{\Lambda Q} - m_\pi$
in the double-differential continuum fragmentation function $\chi_{\Lambda_Q \pi}$,
and thus no feature appears around $m_{\Lambda_Q \pi} \approx m_{\Sigma_Q^*}$
in the double-differential $(m_{\Lambda_Q \pi}, \cos \theta)$
distribution in the background case.
We therefore expect that a standard sideband subtraction
will remain viable to remove the background
also bin by bin in $\theta$.
(Signal-background interference is likewise suppressed
by the narrow widths of the resonances.)
Finally, we stress that care must be taken to avoid biasing
the angular distribution in \eq{decay_distribution_fp}
when applying acceptance cuts on the pion transverse momentum
with respect to the beam axis,
as typically done in spectroscopic analyses~\cite{CDF:2011ac, LHCb:2018haf}
to enrich the sample with signal events
and suppress the $\Lambda_Q \pi$ continuum background.
An attractive way to assess systematic uncertainties
from correcting for the acceptance
(as well as from subleading $1/m_Q$ corrections on the theory side)
is offered by the corresponding angular distribution
for $\Sigma_Q$ baryons,
which is flat in $\cos \theta$ for unpolarized collisions
at leading order in the heavy-quark expansion.

\subsection{Transverse-momentum dependent fragmentation to excited heavy baryons}
\label{sec:decay_distribution_heavy_tmd}

The precise relation of the coefficients $w^{(N,n)}$ in \eq{rho_tmd}
to underlying renormalized nonperturbative matrix elements
differs,
depending on the case considered:
In case~(c) given in the main text, corresponding to heavy-light dihadron
fragmentation, the factorization takes the same multiplicative form
as in \eq{factorization_boosted_heavy_quark_fragmentation}.
For cases~(a) and (b) the matrix elements enter through convolutions
with other TMD matrix elements
encoding the contributions of soft dynamics or other collinear sectors,
none of which however affect the boosted heavy-hadron spin dynamics
or the symmetry analysis leading to \eq{rho_tmd}.
Importantly, the resulting decay pion distribution
for unpolarized collisions
takes a common general form for any of these cases
in terms of the cross section-level spin density in \eq{rho_tmd},
and evaluates to
\begin{align} \label{eq:decay_distribution_tmd}
\frac{1}{\df N_{\Sigma_Q}/\df k_T} \frac{\df N_{\Sigma_Q}}{\df k_T \, \df \cos \theta \, \df \varphi}
&= \frac{1}{4\pi}
\,, \\
\frac{1}{\df N_{\Sigma_Q^{*}}/\df k_T} \frac{\df N_{\Sigma_Q^{*}}}{\df k_T \, \df \cos \theta \, \df \varphi}
&= \frac{1}{4\pi} \biggl\{
   1
   - \frac{3}{8}\bigl[1 + 3 \cos (2\theta)\bigr] \, w^{(2,0)}
   - 3 \cos \theta \, \sin \theta \, \cos \varphi \, w^{(2,1)}
   - \frac{3}{2} \sin^2 \theta \, \cos(2 \varphi) \, w^{(2,2)}
\biggr\}
\nn \,,\end{align}
where $k_T$ is the magnitude of the transverse momentum $\vec{k}_T \perp \vec{z}$,
reconstructed according to the respective experimental scenario.
Importantly,
$\varphi$ in this case is the azimuthal angle of the pion
\emph{relative} to the azimuth of $\vec{k}_T$,
which provides an azimuthal reference direction $\vec{x} = \vec{k}_T/k_T$
in this case, as discussed in the main text.

\end{document}